\documentclass[11pt,a4paper]{article}
\usepackage{epsfig}
\usepackage{wrapfig}

\def\Journal#1#2#3#4{{#1} {\bf #2}, #3 (#4)}
\def\NIMA{{\em Nucl. Instrum. Methods} A}
\def\NPB{{\em Nucl. Phys.} B}

\def\PRL{\em Phys. Rev. Lett.}
\def\PRD{{\em Phys. Rev.} D}

\begin{document}

\begin{center}
{\bf Current Status of Transverse Spin at STAR}

\vspace{0.25in}

Andrew Gordon for the STAR Collaboration

{\em Department of Physics, Brookhaven National Laboratory}

{\em 20 Pennsylvania Street, Upton, NY}

\vspace{0.1in}
Proceedings of the XVIII International Workshop on Deep-Inelastic Scattering and Related Subjects, DIS 2010, April 19-23, 2010, Firenze, Italy

\end{center}

\begin{abstract}
The origins of the proton spin remain an area of active investigation. 
The Relativistic Heavy
Ion Collider (RHIC) at Brookhaven National Lab uniquely
provides polarized proton data at high center of mass energies.
STAR has previously reported large transverse
single spin asymmetries for inclusive neutral pion production
in the forward direction at $\sqrt{s}=200$ GeV, measured
with a modular Pb-glass calorimeter called the ``Forward Pion Detector'' (FPD).
Beginning in 2008 (Run 8), STAR's forward calorimetry was significantly
enhanced
with the commissioning of the ``Forward Meson Spectrometer'' (FMS), which
has full azimuthal coverage in the pseudo-rapidity range $2.5<\eta<4.2$.
This detector has extended asymmetry measurements to higher $P_T$, and its
large areal coverage enables measurements of higher mass
resonances and ``jet-like'' correlations. STAR also provides
excellent tracking and full azimuthal electromagnetic calorimeter coverage at
mid-rapidity, which facilitates analyses of spin-dependent jet structure.
Here we review recent transverse spin measurements with the
STAR detector and also present an update on some existing analyses.
\end{abstract}

\section{Introduction}

STAR has previously reported precision measurements~\cite{PRL2008}
of the analyzing power ($A_N$) of inclusive neutral pions in the forward region.
These measurements used data taken in RHIC runs 3, 5, and 6 with
the Forward Pion Detector (FPD),
a modular Pb-glass array that can be moved horizontally in the
plane transverse to the beamline. These data were taken at
$\sqrt{s}=200$ GeV, where inclusive $\pi^0$ cross sections
are consistent with expectations from pQCD~\cite{PRL2006}. The measurements
showed that the variation of $A_N$ with Feynman-x
($X_F=2P_L/\sqrt{s}$) was qualitatively consistent with expectations from
the Sivers effect~\cite{SIVERS1990},
while the $P_T$ dependence was not. The extent to which the Sivers and
Collins~\cite{COLLINS1993,CH1994}
effects contribute to these measurements has yet to be determined.
FPD measurements of inclusive
$\eta$ production have also been reported, with
the data indicating 
a strong growth of $A_N$ with increasing $X_F$~\cite{HEPPELMANN1}.

\begin{wrapfigure}[26]{l}[0.0in]{8.0cm}
%\begin{figure}[htp]
\includegraphics[width=.5\textwidth]{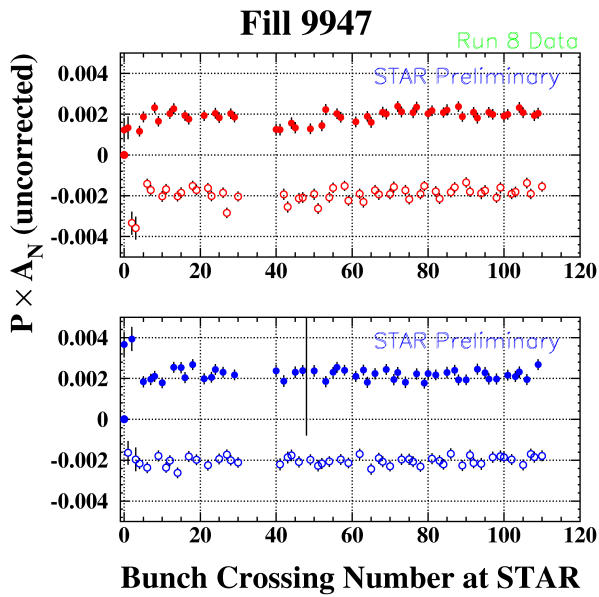}
\caption{\small{Observed transverse single spin asymmetry (uncorrected)
for individual bunches as seen by the STAR BBCs,
for one RHIC fill in Run 8. Top plot is for the beam headed eastward at STAR,
and the bottom is for the beam headed westward.
The solid circles are for bunches that were prepared spin up and the open circles for
spin down. The observed asymmetry was calculated with a procedure similar
to~\cite{KIRYLUK}, and the mean over all the bunches are set to have 0 asymmetry for
each beam.}}
\label{fig:bbc}
%\end{figure}
\end{wrapfigure}

STAR has also observed forward analyzing powers with its ``Beam-Beam-Counters'' (BBC),
at rapidities from $3.9<\eta<5.0$. 
Single transverse spin asymmetries in the
detection of charged hadrons with these detectors can be used as a
``local polarimeter''~\cite{KIRYLUK}. Figure~\ref{fig:bbc} shows the bunch-by-bunch
relative transverse polarization as seen at STAR, for one RHIC fill in 2008 (Run 8).
All fills have
been examined and provide confirmation that the bunches have relatively uniform
absolute polarization, and that the expected bunch-by-bunch spin pattern is also the
spin pattern observed at STAR. The local polarimeter also allows studies of possible
variations in the bunch-by-bunch polarization within a fill.

The Sivers effect identifies the origin of the observed spin asymmetries with
orbital motion of the quarks inside the polarized proton.
%This
%leads to a correlation between the proton spin and the
%intrinsic transverse momentum of the struck quark in the hard
%scatter, which then manifests itself as a Left/Right asymmetry in the resulting
%jet direction.
By contrast, in the Collins
effect the
polarization of the struck quark is correlated to the polarization of the
proton, and the fragmentation of the polarized quark leads to
%Left/Right
asymmetries within the resulting jet.
%It remains to be determined the
%extent to which these two effects contribute to the observed single spin
%asymmetries at STAR.

%\begin{wrapfigure}[21]{l}[0.0in]{8.0cm}
%\begin{figure}[htp]
%\includegraphics[width=.5\textwidth]{An_Pt_2.eps}
%\caption{$A_N$ as a function of $P_{T}(\pi^0)$ for Runs 3, 5, 6 (black circles), Run 8
%with FPD (blue triangles), and Run 8 with FMS (red squares). $X_F$ ranges are indicated
%on the plot. Figure taken from ~\cite{AKIO1}.}
%\label{fig:anpt}
%\end{figure}
%\end{wrapfigure}

STAR commissioned a new detector in 2008 for Run 8,
the Forward Meson Spectrometer (FMS)~\cite{NIKOLA1}.
The FMS is a nearly hermetic array of $1264$ Pb-glass blocks
situated $\sim 700$ cm downstream of the interaction point and spanning
an area $200\times200$ cm$^2$ perpendicular
to the beam pipe. It covers the full azimuth in
the range $2.5<\eta<4.0$ and provides a many-fold increase in the areal
coverage of the forward region at STAR~\cite{NIKOLA1}. This larger detector enhances
the ability to extend the previously published
inclusive $\pi^0$ data~\cite{PRL2008}, and also allows the measurements of higher
mass resonances, including the $\omega$~\cite{GORDON1} and $J/\psi$~\cite{PERKINS1}.

%The large azimuthal coverage of the FMS has allowed single spin asymmetry
%measurements to broken out into azimuthal bins. The data points were fit to the
%function $A_N\cos{\phi}$, and the fitted values of $A_N$ were plotted as a
%function of the $\pi^0$ $X_F$.
The inclusive $\pi^0$ $A_N$ was measured using the FMS in Run 8, and the $X_F$ dependence
of this data has provided an important confirmation of
the previously published
data~\cite{JIM1,AKIO1}. In addition, the FMS has extended STAR's
$P_T$ reach in the forward direction.
%Figure~\ref{fig:anpt} shows $A_N$ measurements as a
%function of $P_T$ of the $\pi^0$.
The Run 8 data agree with the previous data and also indicate that
the analyzing power
in the forward region remains large to at least $P_T>4$ GeV. Moreover, the FMS has
allowed studies of two-$\pi^0$ events. Studies of the azimuthal difference
$\Delta\phi(\pi^0,\pi^0)$ of all
observed $\pi^0\pi^0$ pairs in the FMS show two peaks, one at
$\Delta\phi=0$ and one at 
$\Delta\phi=\pi$,
%indicating a clear jet-like structure to the $\pi^0$ data in the FMS.
indicating that a hard-scatter event underlies these data.

\section{On-going Analyses}

The Collins effect is consistent with string
fragmentation models in which a quark/anti-quark pair is produced
with relative orbital angular momentum at the
point of string breaking~\cite{ARTRU1995,ARTRU1998}.
This leads to an asymmetry within a jet in the
the production of the spin 0 pions, and would lead to the
opposite asymmetry for spin 1 particles~\cite{ARTRU2010}. The decay of the
spin-1, $782$ MeV $\omega$ is accessible with the FMS through
the $\omega\rightarrow\pi^{0}\gamma$ decay channel
(BR=8.9\%~\cite{PDG}), and observation of a negative $A_N$ might provide strong
experimental confirmation of these theoretical models.

The preliminary analysis of $\omega\rightarrow\pi^0\gamma$
decays in the FMS for RHIC Run 8 has been
described elsewhere~\cite{GORDON1}.
Two photon clusters derive from
the $\pi^0$ decay and one directly from the $\omega$. For each set of
three electromagnetic (EM) clusters in the FMS
there are three
possible pairs that can be associated with the $\pi^0$ decay.
The pair whose
mass is closest to $0.135$ GeV/c$^2$ is associated with
the $\pi^0$ decay, while the third is associated with the $\gamma$.
Simulations show that this procedure tags
the photons correctly upwards of $99\%$ of the time.  Minimum energy and $P_T$ cuts
are placed on the clusters to help reduce backgrounds from non-photonic particles.

A signal of $\approx 10$ statistical standard deviations is evident in the data, although
with a signal to background ratio of roughly 1:5~\cite{GORDON1}. This high
background level limits the ability to detect a negative analyzing power.
If we assume that the beam polarization is
$50\%$ and the background has $A_N=0$, then the Run 8 data would be able to detect a
negative $A_N$ with $>2$ statistical standard deviations for $A_N(\omega)<-0.3$.

\begin{wrapfigure}[25]{l}{8.0cm}
%\begin{figure}[htp]
\includegraphics[width=.5\textwidth]{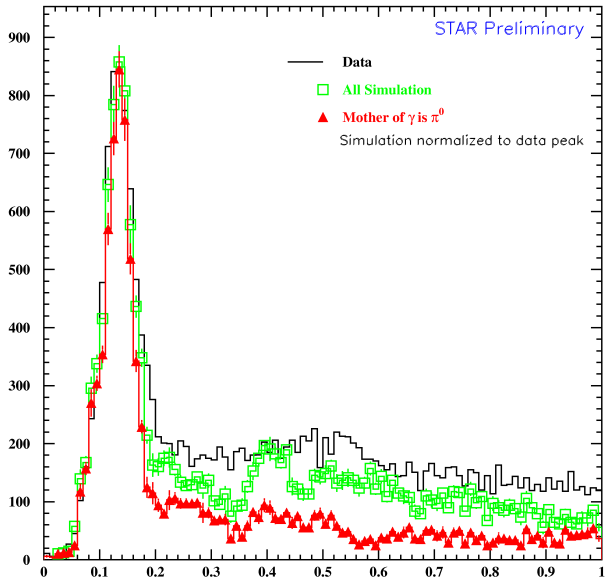}
\caption{\small{Distribution of $M(\gamma,4^{{th}}{cluster})$ (see text) for events
with four or more clusters, for
data (black lines), Pythia(6.2.22)+GEANT simulation (green squares),
and the subset of the simulation for which the mother of the $\omega$-associated photon
is really a $\pi^0$ (red triangles). The green points are
scaled to the data at the peak position.}}
\label{fig:m4pi}
%\end{figure}
\end{wrapfigure}

Pythia(6.2.22)+GEANT simulations \cite{GORDON1} predict that the largest background
source of fake $\omega\rightarrow\pi^0\gamma$ decays are from events where the
$\gamma$ derives from $\pi^0$ decay and not from an $\omega$. There are also
significant contributions from $\eta\rightarrow\gamma\gamma$, as well as from events where the cluster that was assumed to be the $\gamma$ actually
derived from a particle that is not a photon.
The inclusive $\pi^0$ and $\eta$ data sets show large, positive $A_N$ and
also show a strong energy dependence, and the
$\omega$ $A_N$ is thus sensitive to the precise composition of the background.

There is evidence that the Pythia+ GEANT simulation significantly underestimates
the non-photonic background contribution to the three-EM cluster sample. To see this
we look at events that have a set of three clusters that pass the
$\omega$ cuts~\cite{GORDON1}, but that also have at least one extra EM cluster in the
event. We then consider the cluster that we associated with the
$\gamma$ for the $\omega\rightarrow\pi^0\gamma$ decay, and we
calculate the mass of that cluster with one of the extra clusters.
We refer to this
mass as $M(\gamma,4^{{th}}{cluster})$.
We calculate this separately for all extra clusters in the event. 

Figure~\ref{fig:m4pi} shows the distribution of $M(\gamma,4^{{th}}{cluster})$
for data and the Pythia+GEANT simulation.
The peak at
$0.135$ GeV/c$^2$ is from events where the set of three clusters
is actually drawn from the four photons of a $\pi^0\pi^0$ event. The simulation
shows significantly fewer events in the tail of the distribution relative to the
data, when the simulation is scaled so that the peak height matches the data.
This indicates that
the data may contain significantly more non-photonic EM clusters than the simulation.

STAR is also sensitive to the Collins asymmetry through jet reconstruction.
Spin dependence in the
fragmentation of a 
transversely polarized quark in a transversely polarized proton (the Collins effect)
can be manifested in azimuthal dependence of $\pi^{+/-}$ production within the
jet around the jet axis, relative to the direction of the transversely polarized
proton spin~\cite{YUAN1}.
The measured asymmetry depends on $\phi_{\pi}$ and $\phi_S$
where $\phi_{\pi}$ is the azimuth of the pion above the reaction plane,
and $\phi_S$ is the azimuth of the transversely polarized
proton spin relative to the reaction plane.
Here the reaction plane is spanned by a vector along the
transversely polarized beam direction and a vector along the jet axis. 
%The sum is
%over the different transverse spin states $i$ of the proton bunches, with $w_i$
%weights to account for differences in luminosities of the different bunches.

%\begin{wrapfigure}[19]{l}{8.0cm}
%\begin{figure}[htp]
%\includegraphics[width=.4\textwidth]{Fersch_stats.eps}
%\caption{Expected statistical uncertainties (size of error bars, points placed at 0
%on y-axis)
%on Collins jet asymmetries
%as a function of $z$ for $\pi^{+}$, for considering the spin state of the
%``blue'' beam (blue points) and the ``yellow'' beam (yellow points). The EEMC is
%in the forward hemisphere for the blue beam (with no corresponding detector for the
%yellow beam), producing greater statistical power for the blue beam.}
%\label{fig:fersch_stats}
%\end{figure}
%\end{wrapfigure}

The asymmetry can depend on $z=P_{\pi}/P_{jet}$ and also on
$j_{t}$, the component of the pion momentum transverse to the jet axis. The asymmetry
is calculated separately for $\pi^+$ and $\pi^-$.
Since a single spin asymmetry is being measured,
one of the RHIC beams is averaged over,
while the spin state of the other beam
is used for $\phi_S$. Only jets in the forward hemisphere are considered, where
forward is defined relative to the proton beam whose spin state is being considered.
Calculations are done separately for averaging over the spin states of each of the
two beams.

For this analysis, mid-rapidity jets are used. STAR's Time Projection Chamber
(TPC)~\cite{TPC} provides charged particle tracking with good particle identification
for momenta up to $\approx 15$ GeV. In addition the ``barrel''
(BEMC, $-1<\eta<1$)~\cite{BEMC} and ``endcap'' (EEMC, $1<\eta<2$)~\cite{EEMC}
calorimeters provide EM energy measurements and are used for jet reconstruction and also
for triggering.
%The measured asymmetry is
%\begin{equation}
%A=<\sin{(\phi_{\pi}-\phi_S)}>
%\end{equation}

The expected statistical power of transverse data taken in 2006 (Run 6) is
expected to be better than around $\pm 0.01$.
% for %the blue beam data at
%$z$ between $\approx 0.25$ to $0.5$, with the error bars increasing towards higher $z$.
%shown
%in Figure~\ref{fig:fersch_stats}.
By comparison a simulation based on expectations for transversity
and Collins fragmentation from global fits to 
HERMES, COMPASS, and BELLE data~\cite{GLOBAL_1} give a rough expectation for
the asymmetry of $|A|\approx 0.03$.
Systematic uncertainties currently
under study include effects from 
trigger biases, contamination of the $\pi^{+/-}$ sample from
other particles, tracking efficiencies, accuracy of jet and pion kinematic
measurements, beam polarization, and relative luminosities for the different
beam polarization states.

\section{Conclusion and Outlook}

Large single spin asymmetries have been observed at STAR in the forward region, for
both inclusive $\pi^0$ and $\eta$ signals. Run 8 measurements have extended the
$P_T$ range for inclusive $\pi^0$ at high $X_F$ and also confirmed the previous
measurements. In addition, an observation of a negative analyzing power for
$\omega\rightarrow\pi^0\gamma$ in the FMS would provide an exciting test of a
theoretical understanding of the Collins effect. STAR is also analyzing mid-rapidity
jets to provide measurements of the Collins effect.

Future upgrades to the forward region might allow jet triggering, which would allow
direct measurements of the Sivers effect in the forward region. In addition,
STAR has reported measurements of $J/\psi$ production in the forward
region~\cite{PERKINS1}, raising
the question of the degree to which
Drell-Yan production of $e^+e^-$ pairs might also be accessible in the future
in the mass range above $\approx 3$ GeV/c$^2$.

\end{document}